**Self-Trapping Enabled Highly Bright Momentum-Indirect Interlayer Excitons**

Dong Yang[1†], Zisheng Gong[1†], Hao Wen[2], Yue Hu[1], Kaichen Jiang[1], Baixu Xiang[2], Weibo Gao[3,4,5,*], Qihua Xiong[2,6,7,*], Dehui Li[1,8,*]

**Affiliations**

[1]School of Optical and Electronic Information, Huazhong University of Science and Technology, Wuhan 430074, China.

[2]State Key Laboratory of Low-Dimensional Quantum Physics and Department of Physics, Tsinghua University, Beijing 100084, China.

[3]School of Electrical and Electronic Engineering, Nanyang Technological University, Singapore 639798, Singapore.

[4]Division of Physics and Applied Physics, School of Physical and Mathematical Sciences, Nanyang Technological University, Singapore 637371, Singapore.

[5]Centre for Quantum Technologies, Nanyang Technological University, Singapore 639798, Singapore.

[6]Beijing Academy of Quantum Information Sciences, Beijing 100193, China.

[7]Frontier Science Center for Quantum Information, Beijing 100193, China.

[8]Wuhan National Laboratory for Optoelectronics, Huazhong University of Science and Technology, Wuhan 430074, China.

†These authors contributed equally: Dong Yang, Zisheng Gong

*To whom correspondence should be addressed. Emails: wbgao@ntu.edu.sg, qihua_xiong@tsinghua.edu.cn, dehuili@hust.edu.cn

**Abstract**

Interlayer excitons (IXs) in two-dimensional (2D) material heterostructures exhibit large exciton binding energies and long lifetimes, making them ideal platforms for studying excitonic devices and many-body quantum phenomena. However, the spatially separated electron–hole nature of IXs reduces their oscillator strength by two orders of magnitude compared to intralayer excitons. Achieving high-efficiency IX emission remains challenging and requires optimal material selection with appropriate momentum-matching and meticulous device fabrication. Here we demonstrate a highly bright momentum-indirect IX emission within heterostructures formed between 2D perovskites and monolayer transition metal dichalcogenides (TMDs). The quantum yield of IX emission reaches 35.2% on average, over 50 times higher than that of the corresponding constituent TMD monolayer, with the highest value exceeding 60%. Notably, the radiative recombination efficiency of this momentum-indirect IX exceeds that of momentum-direct IXs in monolayer TMD/TMD heterostructures by two orders of magnitude. We suggest that the remarkably bright IX emission in our heterostructure originates from IX self-trapping, induced by strong exciton–phonon coupling arising from the soft lattice nature of the 2D perovskite. Our findings provide new insights into achieving high IX emission efficiency and open new avenues for exploring long-lifetime excitonic devices.

**Main text**

Monolayer transition metal dichalcogenides (TMDs) possess a direct bandgap[1] with strong excitonic effects[2], and high thermal[3] and chemical stability[4]. Stacking two TMD monolayers into a heterostructure typically results in a type-II band alignment, where electrons and holes are spatially separated in different material layers[5]. Due to the spatial confinement effect in two-dimensional (2D) materials, strong Coulomb interactions allow for the formation of interlayer excitons (IXs) with large binding energies[6]. The spatially indirect nature confers IXs longer exciton lifetimes, reaching tens of nanoseconds or even microseconds[7].

However, the spatially separated electron and hole nature of IX makes its emission rather weak[8]. Typically, the oscillator strength of IXs is only two to three orders of magnitude lower than that of the intralayer excitons in TMD/TMD heterostructures[8,9]. In particular, the emission efficiency of momentum-indirect excitons is much weaker than that of direct transitions, making them weakly visible or even dark in spectra in TMD/TMD heterostructures[10]. Therefore, to achieve visible IX emission, it is usually necessary to carefully choose the right TMD materials[11] and stack them with a certain twist-angle to achieve momentum-direct IXs[12]. This requires stringent fabrication conditions, such as h-BN encapsulation[11], annealing[13], and precise twist-angle alignment during the transfer process[12]. From this perspective, it is urgent to explore new material systems or alternative emission routes to achieve efficient IX emission.

Previous studies have demonstrated that robust momentum-indirect IX emission can be achieved in 2D perovskite/monolayer TMD heterostructures regardless of the twist-angle and without requiring a thermal annealing process during the fabrication[14]. IX emission energy can be tuned in a wider energy range from 1.3 to 1.6 eV *via* engineering the layer number of the constituent layer[15] while the IX binding energy can be modulated by changing the organic chains of 2D perovskites[16]. Although substantial progress has been made, the emission efficiency of IXs in these heterostructures and the mechanism underlying such robust momentum-indirect IX emission remain elusive.

Here, we report a highly efficient momentum-indirect IX emission in 2D perovskite/monolayer TMD heterostructures. Regardless of the twist-angle, stable IX emission can be achieved in dry-transferred 2D perovskite/monolayer TMD heterostructures, which is verified by excitation power-dependent photoluminescence (PL), photoluminescence excitation (PLE), time-resolved PL (TRPL), and transient absorption (TA) studies. Remarkably, these IX emissions exhibit an average photoluminescence quantum yield (PLQY) of 35.2%, over 50 times higher than that of the corresponding monolayer TMD, with the highest value exceeding 60%. The radiative recombination efficiency of these momentum-indirect IXs is two orders of magnitude higher than that of the momentum-direct IXs in TMD/TMD heterostructures. Both experiment and calculation reveal a Huang–Rhys factor exceeding 18 for the IX, evidencing strong exciton–phonon coupling[17]. Combined with the soft-lattice nature of the 2D perovskite[18] and ~150 meV broadband IX emission at 77 K, we attribute the anomalously bright momentum-indirect IX emission to self-trapped IXs. In addition, we show that such high-PLQY IX emission can be achieved in diverse combinations of 2D perovskite/TMD monolayer heterostructures, covering a broad emission wavelength range. Our study provides new insights into IX emission, paving the way for future high-quality IX-based light-emitting devices.

**Momentum-indirect IX emission**

Monolayer $MoSe_2$ and $(PEA)_2PbI_4$ (hereafter denoted as PEA, PEA = phenylethylammonium) microplate (~65 nm thick) were first obtained *via* mechanical exfoliation (Fig. S1). A dry transfer technique was then employed to sequentially stack these exfoliated layers onto a 300 nm $SiO_2$/Si substrate to form a heterostructure. Fig. 1a shows the schematic illustration of the device structure, where the monolayer $MoSe_2$ is placed on top of the 2D perovskite microplate, which can protect the 2D perovskite from exposure to oxygen and moisture. Fig. 1b (left) displays optical microscopy images of the device. The electronic band structure of the heterostructure calculated via density functional theory (DFT) is shown in Fig. 1c. The black and red points denote the orbital-projected contributions from monolayer $MoSe_2$ and PEA, respectively. As depicted by the yellow double arrows, the $MoSe_2$/PEA heterostructure exhibits a type-II band alignment, with the valence band maximum originating from the Γ valley of PEA and the conduction band minimum located at the K valley of $MoSe_2$. Therefore, the interlayer transitions from Γ to K are formed.

Fig. 1d displays the normalized PL spectra from the constituent monolayer $MoSe_2$, PEA, and $MoSe_2$/PEA heterostructure regions. In the heterostructure region, a broad emission peak centered around 1.45 eV on the lower energy side of intralayer exciton emission peak of monolayer $MoSe_2$ is observed, which is attributed to IX emission according to previous reports[14], aligning with the calculation result in Fig. 1c. PLE spectrum (Fig. S2) shows that three resonance-enhanced peaks at approximately 1.6 eV, 1.85 eV, and 2.4 eV correspond to the A exciton and B exciton of the monolayer $MoSe_2$ and the free exciton of PEA, respectively. The presence of resonances associated with both constituent layers is a typical feature of IX emission[6]. The PL mapping of IX (~1.45 eV) in Fig. 1b (right) indicates that IX emission is rather uniform across the entire heterostructure region, excluding defect-related localized emission as its origin. The excitation power-dependent PL spectra show that the IX emission peak undergoes a continuous blueshift, which is attributed to the dipole-dipole repulsion resulting from the permanent vertical dipole moment of the IX[19] (Fig. S3). Additionally, the TRPL measurements reveal a significantly long PL decay time of up to 43 ± 2 ns, which is typical for IXs[7] (Fig. 1e).

TA measurements at 77 K further confirm the transfer of photoexcited holes from $MoSe_2$ to PEA. Under 633 nm pumping, a PEA ground-state bleaching at ~2.4 eV appears exclusively in the heterostructure region, but is absent in monolayer $MoSe_2$ (Fig. 1f and Fig. S4). As the pump energy is ~400 meV below the PEA bandgap, only $MoSe_2$ is directly excited and the PEA bleaching observed in the heterostructure can only arise from the transfer of photogenerated holes. Furthermore, when probing the $MoSe_2$ A exciton, the heterostructure exhibits a faster initial decay, consistent with ultrafast hole transfer from $MoSe_2$ to PEA driven by the type-II alignment[20] (Fig. 1g). Detailed TA results and analyses are provided in the Supplementary Note 1. Combined with the PL characteristics, the TA results further confirm that the broad emission peak originates from IXs.

**High PLQY of IX emission**

Under identical laser excitation power and PL collection conditions, we compared the emission intensity of IXs within the heterostructure region to that of intralayer excitons in the constituent monolayer $MoSe_2$ region within the same device. The IX emission is significantly stronger and has a broader full width at half maximum (FWHM), as shown in Fig. 2a. The IX peak (blue) exhibits a FWHM of approximately 150 meV, whereas intralayer excitons ($X_A$), including the neutral exciton ($X^0$, orange) and charged exciton ($X^-$, pink), display a narrower FWHM of ~20 meV. In addition, as the excitation power increases by nearly three orders of magnitude, the IX emission peak exhibits a continuous blueshift while maintaining its broadband emission (Fig. 2b). The much broader and stronger IX emission peak compared to that of intralayer excitons heralds a higher luminescence efficiency.

We calculated excitation power-dependent PLQY of IX emission and compared with that of $X_A$ ($X^0$ + $X^-$) emission in monolayer $MoSe_2$ (Fig. 2c). Details of the PLQY calculation are provided in Supplementary Note 2. The PLQY of monolayer $MoSe_2$ is below 1%, consistent with the previous reports[21,22]. In contrast, the PLQY of IX emission is much higher, reaching ~38% at low excitation power. In addition, the PLQY of IX emission decreases faster than that of intralayer exciton with increasing excitation power, which can be readily understood by considering the much longer lifetime of IXs, making them more susceptible to high-density accumulation that induces nonradiative processes such as exciton–exciton annihilation in a 2D confined system[23].

Furthermore, we fabricated another trilayer device to directly compare the emission efficiency of momentum-indirect IXs in the $MoSe_2$/PEA heterostructure and momentum-direct IXs in TMD/TMD heterostructures. An additional monolayer $WSe_2$ was stacked onto the $MoSe_2$/PEA heterostructure with precise long-edge alignment with $MoSe_2$. This alignment of the trilayer heterostructure allows us to simultaneously observe the momentum-indirect IX emission in the $MoSe_2$/PEA heterostructure (denoted as $IX_{T/P}$) and the momentum-direct IX emission in the $WSe_2/MoSe_2$ heterostructure[24] (denoted as $IX_{T/T}$) within the same device. As shown in Fig. 2d, in the $WSe_2/MoSe_2$ heterostructure region, an emission peak around 1.38 eV was observed, which is attributed to momentum-direct $IX_{T/T}$ according to the extensive previous studies[24,25]. In the $MoSe_2$/PEA heterostructure region, only the momentum-indirect $IX_{T/P}$ emission peak is present. In the $WSe_2/MoSe_2$/PEA trilayer heterostructure region, emission peaks from both $IX_{T/T}$ and $IX_{T/P}$ are observed. Remarkably, the PL intensity of the momentum-indirect $IX_{T/P}$ is significantly stronger than that of the momentum-direct $IX_{T/T}$. The PL mappings of this device reveal that the IX emission is spatially rather homogeneous across the respective regions (Fig. S7). In addition, strong IX emission dominates the PL spectra of the $WSe_2/MoSe_2$ region, suggesting efficient interlayer coupling and favorable momentum alignment between $MoSe_2$ and $WSe_2$ layers[10]. IX emission peaks in all three regions show a blueshift and a sublinear increase with increasing excitation power, agreeing well with the behavior of the IX emission[6] (Fig. S8).

To quantitatively evaluate the emission efficiency of the momentum-indirect $IX_{T/P}$, we calculated the ratio of the radiative recombination efficiencies between $IX_{T/P}$ and $IX_{T/T}$. Here, we approximate the decrease of $IX_{T/P}$ emission intensity in the $WSe_2/MoSe_2$/PEA heterostructure, relative to that in the $MoSe_2$/PEA heterostructure, by considering two effects: the reduced absorption of $MoSe_2$ due to the top monolayer $WSe_2$ and the competing charge transfer from $MoSe_2$ to PEA and $WSe_2$. The PL of the IXs can be well fit by a single peak, and the typical Voigt fitting results are shown in Fig. S9. Combining the emission intensities of the individual IXs in the $WSe_2/MoSe_2$ and $MoSe_2$/PEA heterostructures, we obtained the excitation-power-dependent effective radiative efficiency ratio between $IX_{T/P}$ and $IX_{T/T}$, shown in Fig. 2e (detailed calculations are provided in Supplementary Note 3). From this analysis, we find that the effective radiative efficiency ratio of $IX_{T/P}$ to $IX_{T/T}$ reaches a maximum of approximately 300, indicating that the radiative recombination efficiency of the momentum-indirect $IX_{T/P}$ is significantly higher than that of the momentum-direct $IX_{T/T}$. For another device shown in Fig. S10, this ratio reaches a maximum of approximately 160. In this case, only the $IX_{T/P}$ emission with a relatively reduced intensity was observed in the $WSe_2/MoSe_2$/PEA heterostructure, and no detectable emission was present from $IX_{T/T}$.

We statistically analyzed all fabricated devices, including 34 $MoSe_2$/PEA heterostructures and 27 $WSe_2/MoSe_2$ heterostructures, and compared the corresponding PLQY of IXs under ~1 μW excitation, together with that of monolayer $MoSe_2$, as shown in Fig. 2f. We find that the average PLQY of IX emission in $MoSe_2$/PEA heterostructures reaches 35.2%, representing an enhancement of approximately two orders of magnitude compared to IXs in $WSe_2/MoSe_2$ (~0.5%) and intralayer $X_A$ emission in monolayer $MoSe_2$ (~0.6%). It should be noted that the majority of $WSe_2/MoSe_2$ heterostructures exhibit extremely weak IX emission that is not detectable under ~1 μW excitation; therefore, the reported average PLQY of 0.5% is obtained from the nine devices with detectable IX emission. Meanwhile, the statistics in Fig. 2f further show that, despite the additional variability inevitably introduced by heterostructure fabrication, the IX PLQYs in $MoSe_2$/PEA heterostructures remain consistently high and exhibit a narrower sample-to-sample distribution on the logarithmic scale than those of monolayer $MoSe_2$. This suggests the robust and reproducible nature of the bright IX emission in the $MoSe_2$/PEA heterostructure. Remarkably, among the 34 $MoSe_2$/PEA heterostructures measured, 13 devices exhibit PLQYs exceeding 40%, with the highest value reaching 65.3%. The PL spectrum of this highest-PLQY device is shown in Fig. S11. Such a high PLQY likely arises from the combined effects of a clean heterointerface and robust interfacial coupling.

**Self-trapped IXs**

The anomalously high PLQY and broadband characteristics of the momentum-indirect IX emission imply that IXs in our heterostructures might be self-trapped[26]. In semiconductors with sufficiently strong exciton–phonon coupling, excitons can become localized even in defect-free and undisturbed crystal lattices[27]. This phenomenon is referred to as exciton self-trapping, and the radiative recombination of self-trapped excitons is an intrinsic luminescence mechanism typically accompanied by broadband emission[26]. Exciton self-trapping has therefore been widely exploited to enhance the quantum yield of semiconductors[28-30]. The self-trapped states can strongly suppress nonradiative recombination pathways, thereby significantly enhancing the PLQY[30]. More importantly, the formation of self-trapped excitons intrinsically mixes electronic and vibrational states and localizes excitons in real space, thereby effectively relaxing the momentum-conservation constraints for radiative recombination[27]. As a result, self-trapped excitons can exhibit broadband and highly efficient radiative emission even in systems where conventional excitonic transitions are optically forbidden[28].

Fig. 3a illustrates the proposed emission mechanism of self-trapped IXs in our $MoSe_2$/PEA heterostructure. Optical excitation first generates spatially separated IXs, which subsequently become localized within a potential well created by strong lattice deformation over the surrounding crystal lattice, particularly within the soft lattice of the 2D perovskite[18] hosting the hole. Such real-space localization of the self-trapped IX inevitably introduces a strong delocalization of the wavefunction in momentum-space[31], enabling quasi-direct radiative recombination of our momentum-indirect IX. Fig. 3b schematically illustrates this process in the corresponding energy band diagram. This self-trapping IX scenario provides a plausible explanation for the unusually high PLQY observed in our momentum-indirect IXs.

The restricted open-shell Kohn–Sham (ROKS) theory was then employed to theoretically uncover the origin of the broadband emission[32]. Fig. 3c shows the calculated electron and hole density distributions in the excitonic polaron state for IXs in the $WSe_2/MoSe_2$ (left) and $MoSe_2$/PEA (right) heterostructures, respectively. It is evident that the IXs in the $MoSe_2$/PEA heterostructure are much more localized in real space, indicating the formation of self-trapped IXs, while the IXs in the $WSe_2/MoSe_2$ heterostructure behave more like delocalized free excitons. Fig. 3d presents the configuration coordinate diagram of the IX self-trapping process in $MoSe_2$/PEA heterostructures. The self-trapping energy $E_{st}$ and lattice-deformation energy $E_d$ were calculated to be 70.9 meV and 401 meV, respectively, corresponding to the energy differences between the self-trapped and free-exciton configurations in the excited and ground states. Calculation details are provided in Supplementary Note 4. Most importantly, a Huang–Rhys factor of $S$ = 18.3 confirms the strong exciton–phonon coupling in the $MoSe_2$/PEA heterostructure[17]. The Huang–Rhys factor is a dimensionless quantity that characterizes exciton-phonon coupling and serves as a key descriptor of the strength of exciton-phonon interaction[33]. Moreover, this Huang–Rhys factor is favorable for efficient broadband self-trapped exciton emission, whereas excessively large $S$ values can induce nonradiative transitions and reduce the PLQY[34].

Experimentally, the exciton–phonon coupling strength is closely related to the luminescence linewidth and it can be quantified by fitting the temperature-dependent phonon-assisted PL broadening using the following equation[29]:

$$\mathrm{FWHM} = 2.36\sqrt{S}\,\langle\hbar\omega\rangle\sqrt{\coth\left(\frac{\langle\hbar\omega\rangle}{2k_\mathrm{B}T}\right)}$$

Here, $T$ and $k_B$ denote the temperature and Boltzmann constant, respectively. The fitting parameters $S$ and $\langle\hbar\omega\rangle$ correspond to the Huang–Rhys factor and the average phonon energy, respectively. To this end, we extracted the temperature-dependent FWHM broadening of IXs in the $MoSe_2$/PEA and $WSe_2/MoSe_2$ heterostructures from Fig. S12, measured on the device shown in Fig. 2d. The fitting in Fig. 3e yields an effective Huang–Rhys factor of 18.1 for IXs in the $MoSe_2$/PEA heterostructure, consistent with the calculated result, further confirming the strong exciton–phonon coupling. In contrast, the same fitting of IXs in the $WSe_2/MoSe_2$ heterostructure gives a much smaller Huang–Rhys factor of 0.4, consistent with previously reported values and validating its weak exciton–phonon coupling nature[35,36]. Notably, the average phonon energy obtained from the fitting for IXs in the $MoSe_2$/PEA heterostructure is 11.3 meV (91 $cm^{-1}$), matching the inorganic octahedral distortion modes of PEA, particularly the out-of-plane Pb–I bond stretching mode[37], whereas that for IXs in the $WSe_2/MoSe_2$ heterostructure is 29.4 meV (237 $cm^{-1}$), more consistent with the out-of-plane optical phonon modes of the TMDs[12]. This consistency indicates that the strong exciton–phonon coupling of the $IX_{T/P}$ originates from the rich phonon modes of the 2D perovskite[18], consistent with our proposed mechanism.

Taken together, both the theoretical calculations and experimental observations consistently support the self-trapped IX picture in the $MoSe_2$/PEA heterostructure, where strong exciton–phonon coupling induced by the soft 2D perovskite lattice[18] enables efficient broadband radiative emission from otherwise momentum-indirect IXs.

**Universality of self-trapped IXs in TMD/2D perovskite heterostructures**

One of the key advantages of IXs lies in their compatibility with band-gap engineering through heterostructure design[38]. Motivated by this, we fabricated a series of heterostructures combining representative TMD monolayers with different 2D perovskites. We find that the emission of momentum-indirect IXs is broadly observed across various TMD and 2D perovskite combinations, including $WSe_2$/EA, $WSe_2$/BA, $WSe_2$/HA, $WSe_2$/PEA, $MoSe_2$/BA and $WS_2$/PEA, as shown in Fig. 4a under a fixed excitation power of ~1 μW. For convenience, we denote the linear-chain 2D perovskites with the general formula $(C_nH_{2n+1}NH_3)_2PbI_4$ (n = 2, 4, 6) as EA, BA, and HA, respectively. Similarly, high PLQY IX emission is consistently observed across all heterostructures, with PLQYs exceeding those of intralayer $X_A$ emission from the corresponding monolayer TMDs, as summarized in Fig. 4b (10.3–43.4% versus ≤0.5%). While the IX emissions span different energy ranges, they all exhibit similar broadband emission characteristics, indicating the formation of self-trapped IXs in all these heterostructures. Although it is impractical to explore all possible material combinations, our results nevertheless suggest that high PLQY IX emission formed between 2D perovskites and TMD monolayers is universal. The broad spectral coverage of the IX emission further highlights the potential of these systems for tunable excitonic devices.

**Discussion**

Our findings, to some extent, overcome the limitations of conventional TMD/TMD heterostructures, where low-oscillator-strength IX emission typically requires careful momentum matching and complex heterostructure design[8,12]. By significantly enhancing the PLQY of momentum-indirect IXs, we demonstrate a new material platform and radiative mechanism for IXs. Our results offer new insights into the study of excitonic many-body effects, and open up opportunities for the development of high-performance exciton-based optoelectronic devices. Beyond those, our work motivates the exploration of new soft-lattice hybrid heterostructures guided by the self-trapped IX mechanism, potentially opening a route toward high-PLQY IX emission.

**Methods**

**Device fabrication.** Monolayer $MoSe_2$, few-layer BN, and few-layer graphene were mechanically exfoliated from bulk crystals purchased from HQ Graphene. 2D perovskite crystals were synthesized using a previously reported solution-based method[39]. 2D perovskite nanosheets with desired thicknesses were then exfoliated from the bulk crystals. All materials were sequentially dry-transferred onto wafers using polydimethylsiloxane (PDMS) stamps.

**Optical characterizations.** Optical images were captured using an Olympus BX53M optical microscope. PL and Raman spectra were measured with a home-built Raman spectrometer (Horiba, iHR-550) equipped with a liquid-nitrogen-cooled charge-coupled device (CCD). The samples were put in a cryostat with a liquid nitrogen bath (Cryo Industries of America Inc.) equipped with a temperature controller (Lake Shore Cryotronics, Model 336). For the perovskite crystals, excitation was provided by a 405 nm solid-state laser. TMDs and heterostructures were excited using a 633 nm He-Ne laser. Unless otherwise noted, the PL spectra in this study were measured at 77 K using 633 nm laser excitation, focused by a ×50 objective (NA = 0.6, Nikon) with a spot size of approximately 1.2 μm. TRPL measurements were conducted using a 640 nm picosecond laser for excitation (PicoQuant, Prima). The detection system consisted of a TCSPC system (PicoQuant, TimeHarp 260) coupled to the spectrometer. For PLE experiments, the samples were excited by a pulsed laser (78.1 MHz) generated by a supercontinuum light source (NKT Photonics, SuperK Fianium FIU-15) with an acousto-optic tunable filter (YSL Photonics, AOTF). TA measurements were performed using a commercial pump–probe spectrometer (Helios Fire, Ultrafast Systems). 800 nm pulses (35 fs, 1 kHz) from a Ti:sapphire amplifier (Coherent Inc.) were split into pump and probe paths. The pump beam was directed into an optical parametric amplifier to generate tunable excitation pulses. The spectral bandwidth of the pump pulses was further defined using long- and short-pass filters. A white-light continuum probe was produced by focusing the second 800 nm beam into a sapphire plate and temporally delayed using a motorized delay stage. The pump and probe beams were spatially overlapped on the sample with spot diameters of ~200 μm and ~6 μm, respectively. The transmitted probe was detected by a spectrometer and a back-thinned CCD array. The differential transmission signal was obtained as $\Delta T/T = (T_{on} - T_{off}) / T_{off}$. The hBN/$MoSe_2$/PEA heterostructure was dry-transferred onto a fused quartz substrate for TA measurements.

**First-principles density functional theory calculations.** Based on density functional theory (DFT), the first-principles calculations were performed using the Vienna Ab initio Simulation Package (VASP)[40]. The exchange-correlation effects were treated within the framework of the Perdew–Burke–Ernzerhof (PBE) generalized gradient approximation (GGA)[41]. The DFT-D3 method was employed to correct for van der Waals (vdW) interactions. A plane-wave energy cutoff of 500 eV was used, and a 3×3×1 Γ-centered k-point grid was adopted for self-consistent calculations. For ionic optimization, a 1×1×1 k-grid was applied. Dipole corrections were included to eliminate dipole moments arising from the asymmetry of the upper and lower surfaces of the heterostructure. Additionally, a 20 Å vacuum layer was introduced along the z-direction to prevent interlayer interactions[42]. All structures were fully relaxed until the system energy converged to $10^{-5}$ eV, and the maximum ionic forces were below a threshold of 0.02 eV/Å. The lattice constants of monolayer $MoSe_2$ are a = b = 3.317 Å, and those of 2D perovskite are a = 8.606 Å and b = 8.474 Å. After the heterostructure is formed, the lattice constants of the heterostructure are a = 19.853 Å and b = 8.738 Å. Each supercell of the heterostructure is composed of a [[1,2], [1,0]] unit cell of 2D perovskite and a [[4,7], [3,1]] unit cell of monolayer $MoSe_2$. The lattice mismatch ratio of the heterostructure is 3.6%. The interlayer spacing of the heterostructure is 2.72 Å.

**Acknowledgments**

This work is supported by National Key Research and Development Program of China (2024YFA1208500), Fundamental and Interdisciplinary Disciplines Breakthrough Plan of the Ministry of Education of China (JYB2025XDXM121), Hubei Provincial Natural Science Foundation (2025AFA039), Open Project Program of Hubei Optical Fundamental Research Center (HBO2026C015) and TCL Science and Technology Innovation Fund. We thank the Analytical and Testing Center of Huazhong University of Science and Technology for support of the perovskite characterizations and thank the Center of Micro-Fabrication of WNLO for support of the device fabrication.

**Author contributions:**

D. Y. and D. L. conceived the initial idea. D. Y., W. G., Q. X., and D. L. designed the experiments. D. Y. and Z. G. fabricated the devices. Y. H., K. J., and D. L. measured the collection efficiency of the home-built PL measurement system. Z. G., H. W., Q. X., and B. X. performed the TA measurements. D. Y. carried out the other optical measurements. Z. G. performed the theoretical calculations. D. Y., Z. G., W. G., Q. X., and D. L. wrote the manuscript with contributions from all authors. All authors contributed to the discussion of the results and the revision of the manuscript.

**Competing Interests Statement:**

The authors declare no competing interests.

**Data availability:**

All data supporting the findings of this study are available within the paper and the Supplementary Materials. Additional data are available from the corresponding authors upon reasonable request.

**Figure legends**

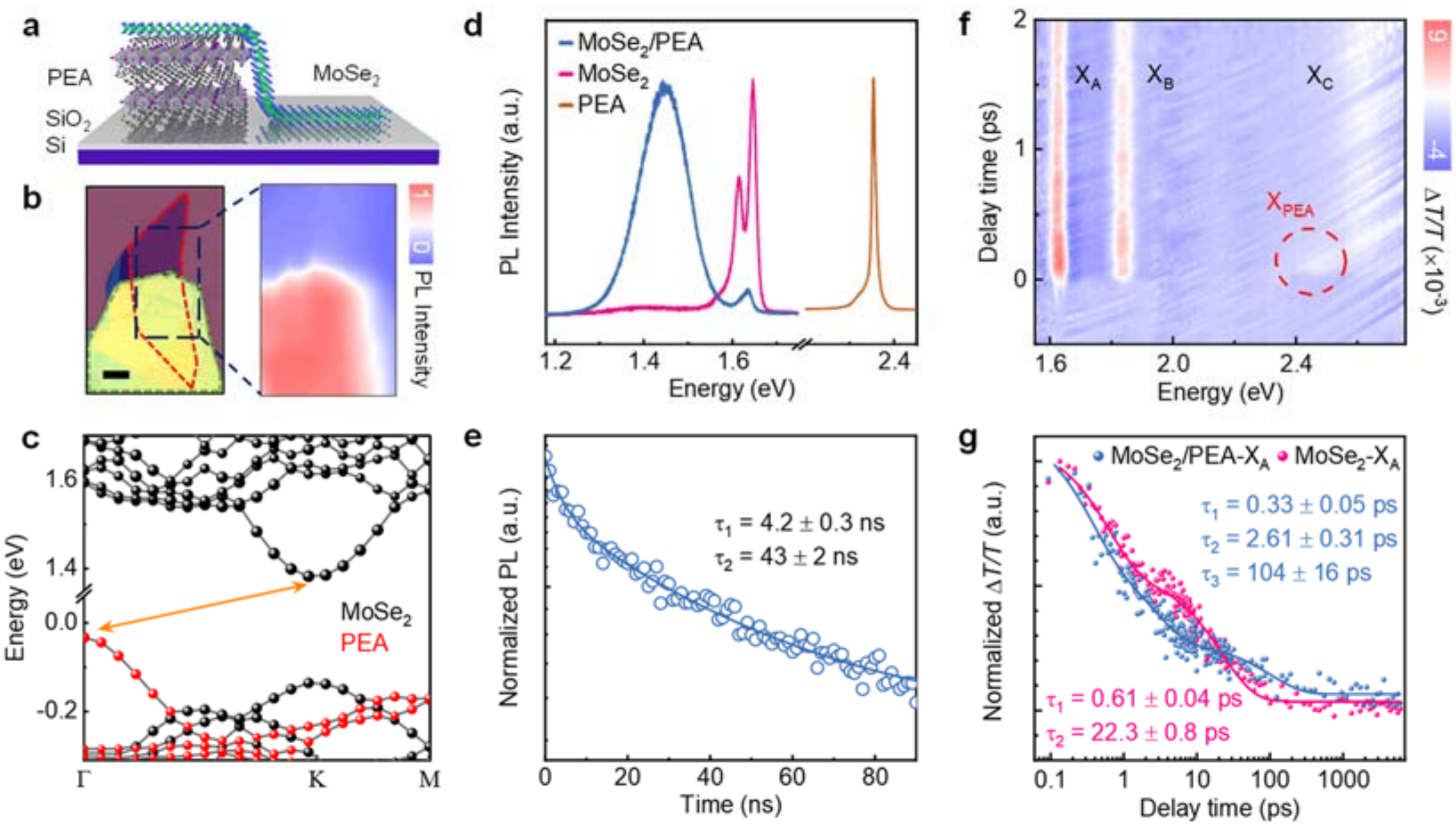

**Fig. 1 | Momentum-indirect IX emission. a,** Schematic illustration of the $MoSe_2$/PEA heterostructure. **b**, Optical micrograph of the as-fabricated sample. Red and green dashed lines outline the monolayer $MoSe_2$ and the PEA microplate, respectively. The blue rectangle marks the PL mapping region for the IX emission at ~1.45 eV. Scale bar: 10 μm. **c**, Computed band structure of the $MoSe_2$/PEA heterostructure. The yellow double-headed arrow indicates Γ−K interlayer transition. **d**, Normalized PL spectra of monolayer $MoSe_2$, PEA, and heterostructure regions. **e**, TRPL dynamics of the IX. The solid line is the double-exponential fitting result. **f**, 2D $\Delta T/T$ map as a function of probe energy and delay time under 633 nm excitation for the $MoSe_2$/PEA heterostructure. **g**, TA dynamics of the $MoSe_2$ A exciton extracted from monolayer $MoSe_2$ and the $MoSe_2$/PEA heterostructure. The solid lines represent the exponential fitting curves.

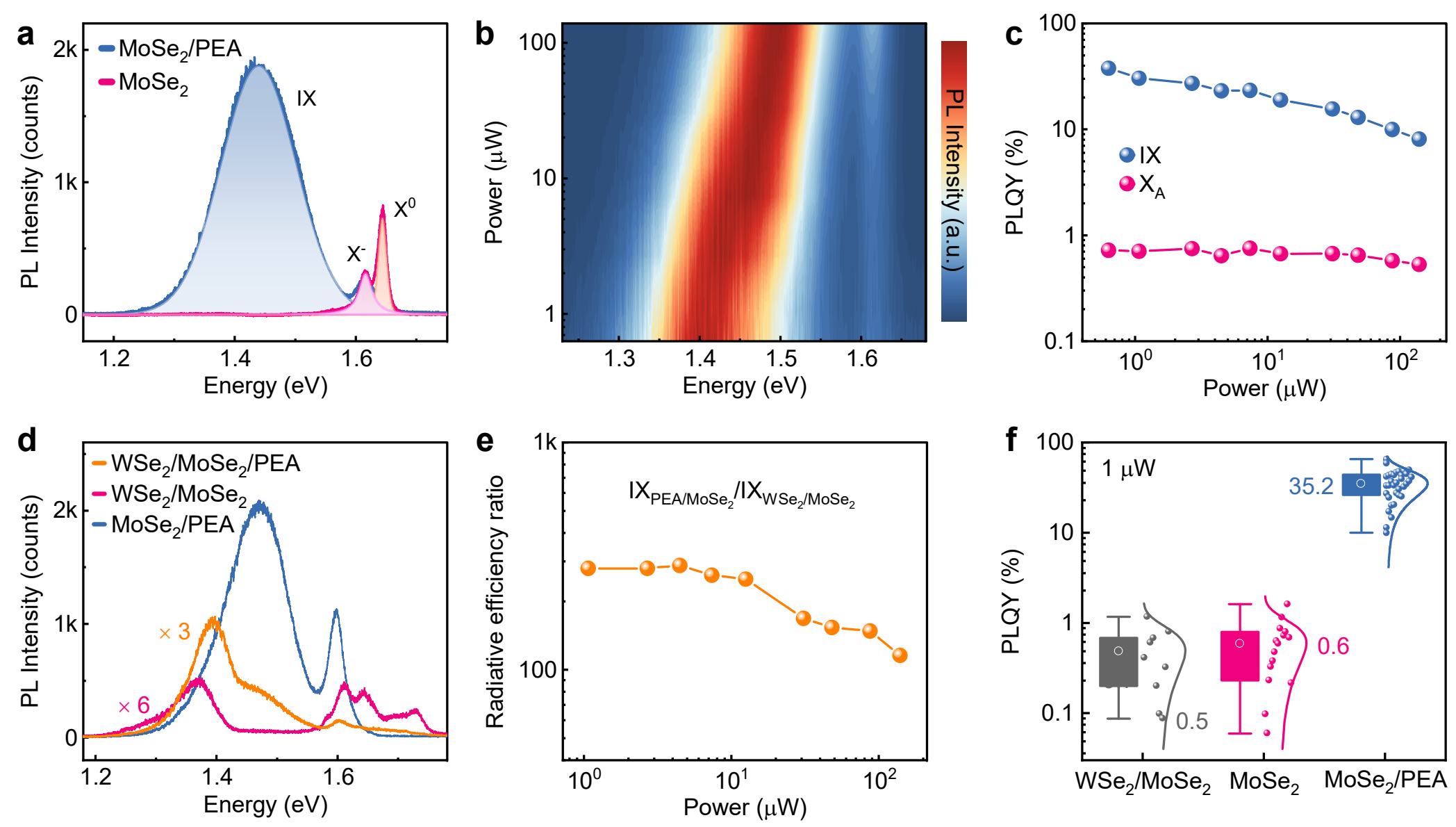

**Fig. 2 | High PLQY of IX emission. a,** PL spectra of the $MoSe_2$/PEA heterostructure (blue) and the monolayer $MoSe_2$ (red) in a $MoSe_2$/PEA heterostructure under 4.47 μW excitation. Shadows represent fitted IX (blue), $X^-$ (pink), and $X^0$ (orange) emission peaks. **b**, Normalized excitation power-dependent PL of the heterostructure. **c**, Power-dependent PLQY of IX (blue) and $X_A$ (red). **d**, PL spectra of the $WSe_2/MoSe_2$ heterostructure (red), $WSe_2/MoSe_2$/PEA heterostructure (orange) and $MoSe_2$/PEA heterostructure (blue) under 12.5 μW excitation. The spectra of the $WSe_2/MoSe_2$/PEA and $WSe_2/MoSe_2$ heterostructures are multiplied by factors of 3 and 6, respectively, for clarity. **e**, Effective radiative efficiency ratio between $IX_{T/P}$ and $IX_{T/T}$ as a function of excitation power. **f**, Statistical comparison of the PLQY measured from IXs in $MoSe_2$/PEA heterostructures, $WSe_2/MoSe_2$ heterostructures, and $X_A$ in monolayer $MoSe_2$ under ~1 μW excitation. The labeled values indicate the average PLQY of each group.

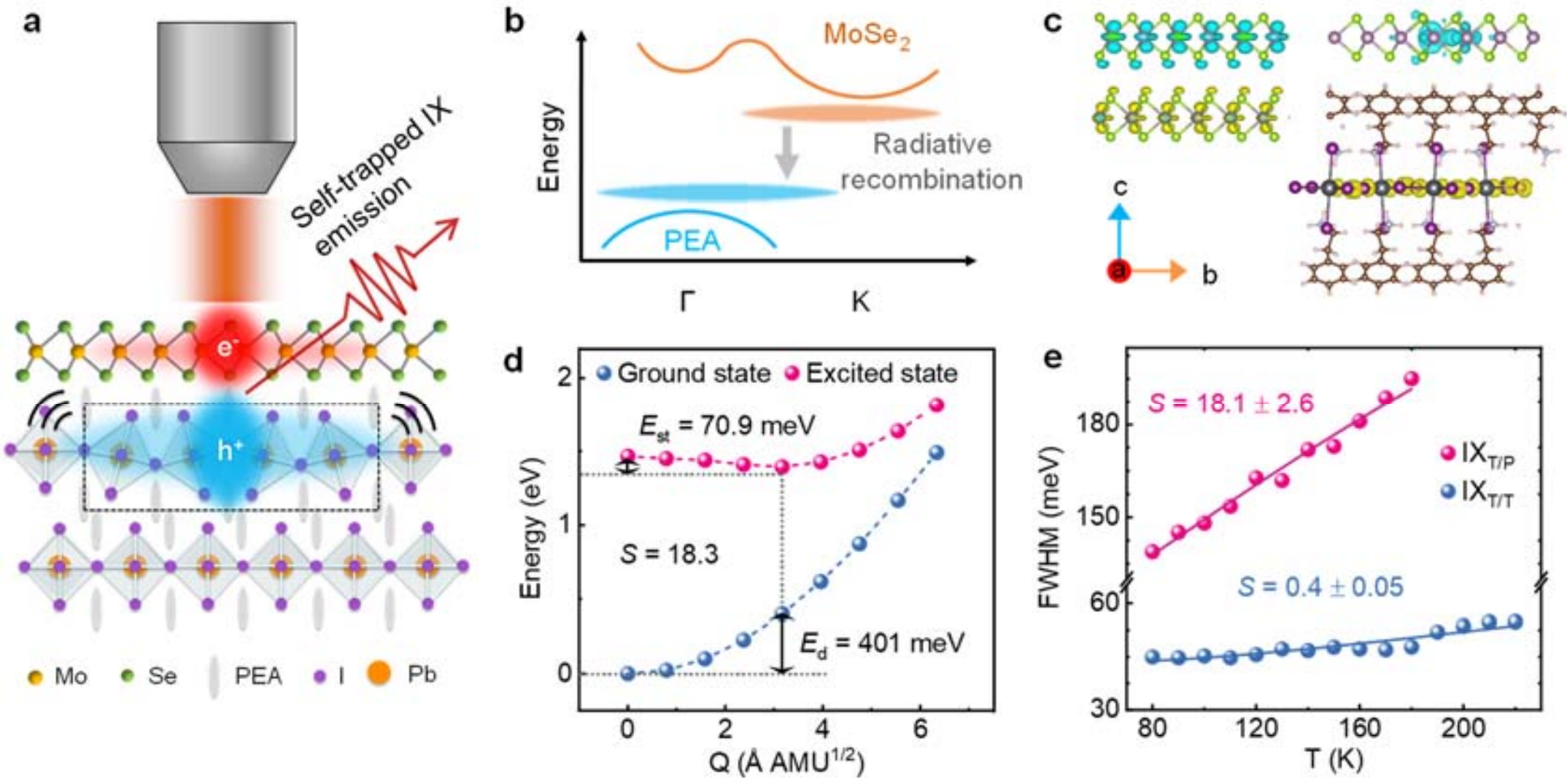

**Fig. 3 | Self-trapped IXs. a,** Schematic illustration of the self-trapped IX emission mechanism in the $MoSe_2$/PEA heterostructure. **b,** Schematic illustration of the self-trapped IX in momentum space. Orange (blue) filled areas denote the amplitude of the electron (hole) wavefunction. The gray arrow indicates quasi-direct radiative recombination. **c,** Calculated charge densities of electron (cyan) and hole (yellow) wavefunctions for IXs in the $WSe_2$/$MoSe_2$ heterostructure (left) and the $MoSe_2$/PEA heterostructure (right). **d,** Configuration coordinate diagram of the IX self-trapping process in the $MoSe_2$/PEA heterostructure. $E_{st}$, $E_d$, and $S$ denote the self-trapping energy, lattice-deformation energy, and Huang–Rhys factor, respectively. **e,** Temperature-dependent PL linewidth broadening of IXs in the $MoSe_2$/PEA and $WSe_2$/$MoSe_2$ heterostructures. Solid curves are fits using the phonon-assisted PL broadening model.

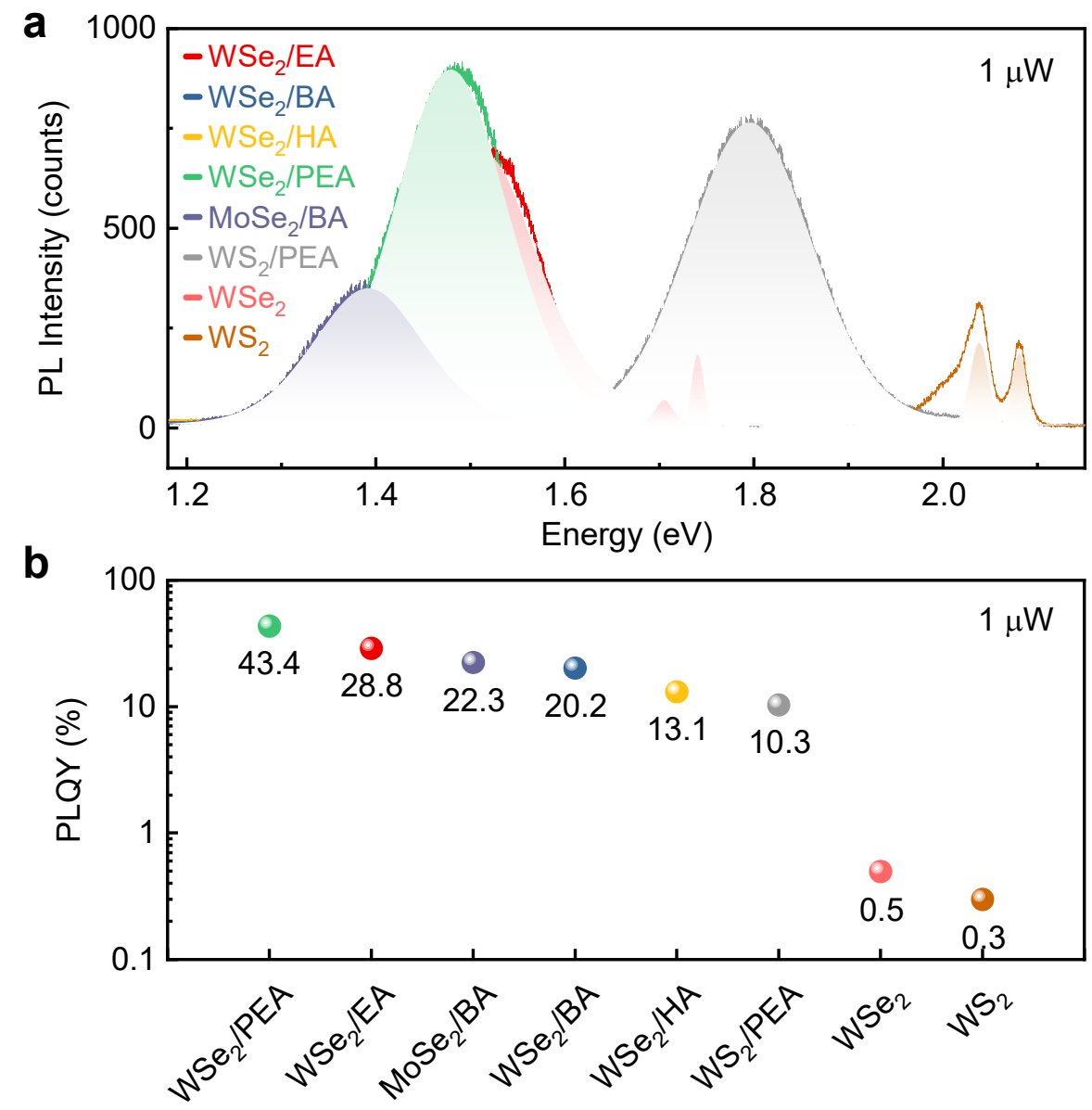

**Fig. 4 | Universality of self-trapped IXs in TMD/2D perovskite heterostructures. a**, PL spectra of representative TMD/2D perovskite heterostructures measured under a fixed excitation power of ~1 μW, including $WSe_2$/EA, $WSe_2$/BA, $WSe_2$/HA, $WSe_2$/PEA, $MoSe_2$/BA, and $WS_2$/PEA heterostructures, together with monolayer $WSe_2$ and $WS_2$ for comparison. **b**, PLQY extracted from the corresponding PL spectra in Fig. 4a for IX emissions in different TMD/2D perovskite heterostructures and $X_A$ emissions in monolayer TMDs under ~1 μW excitation.